\documentclass[12pt]{article}
\usepackage{epsfig,amsfonts,amssymb}
\usepackage{hyperref, caption}
\usepackage{subfig}
\usepackage{graphicx}% Include figure files
\usepackage{dcolumn}% Align table columns on decimal point
\usepackage{bm}% bold math
\usepackage{slashbox,amsmath}
\numberwithin{equation}{section}
\usepackage{setspace}
\usepackage{epsfig}
\usepackage{etoolbox}
\apptocmd{\sloppy}{\hbadness 10000\relax}{}{}
\input epsf.sty
\usepackage{cite}
\newcommand{\nn}{\nonumber}
\newcommand{\be}{\begin{equation}}
\newcommand{\ee}{\end{equation}}
\newcommand{\bea}{\begin{eqnarray}}
\newcommand{\eea}{\end{eqnarray}}
\usepackage[T1]{fontenc}
\usepackage[utf8]{inputenc}
\usepackage{multirow}
\usepackage{array}
\usepackage{wrapfig} 
\usepackage{tabularx}
\usepackage{setspace}
\usepackage{slashed}

\begin{document}
\pagenumbering{gobble} 
\baselineskip 24pt

\begin{center}
{\Large \bf Study of confinement/deconfinement transition  in  AdS/QCD with generalized warp factors }

\end{center}

\vskip .6cm
\medskip

\vspace*{4.0ex}

\baselineskip=18pt

\centerline{\large \rm Shobhit
Sachan%$^\dagger$ and Sanjay Siwach$^*$ 
}

\vspace*{4.0ex}

\centerline{\large \it Department of Physics,}
\centerline{\large \it  Banaras Hindu University}
\centerline{\large \it  Varanasi, 221005, India}

\vspace*{1.0ex}
\centerline{\small E-mail: 
shobhitsachan@gmail.com}

\vspace*{5.0ex}

\centerline{\bf Abstract} \bigskip
\noindent
We study analytical solutions of charged black holes and thermally charged AdS with generalized warped factors in Einstein-Maxwell-Dilaton system. We calculate Euclidean action for charged AdS and  thermally charged AdS. The actions in both backgrounds are regularized by the method of background subtraction. The study of phase transition between charged black hole and thermally charged AdS gives an insight to the confinement/deconfinement transition. The plots of grand potential vs temperature and chemical potential  vs transition temperature are obtained.

\vspace*{3.0 ex}
\vfill \eject

\baselineskip=18pt
%%%%%%%%%%%%%%%%%%%%%%%%%%%%%%%%%%%%%%%%%%%%%%%%%%%%%%%%%%%%%%%%%%%%%%%%%%%%%%%%%%%%%%%%%
\newpage 
\pagenumbering{arabic} 
\section{Introduction}
Strongly interacting systems are  always a challenge to our analytical knowledge. Quantum Chromodynamics is such a theory, which can not be solved analytically in low energy regime. There are two methods two solve QCD; one is `lattice QCD' \cite{Kogut:2004ca} and other is `AdS/QCD'. The formulation of lattice QCD is based on discretization of space-time and it  requires high performance computing. On the other hand, AdS/QCD is analytic approach and motivated by the gauge/gravity duality \cite{Maldacena:1997re, Witten:1998qj,Gubser:1998bc,Witten:1998zw}.  Some properties of QCD like theories motivated by gauge/gravity duality such as confinement and chiral symmetry breaking have been studied extensively in \cite{Brandhuber:1998er, Erlich:2005qh,
Karch:2006pv, Lee:2009bya, Herzog:2006ra, BallonBayona:2007vp, Megias:2010ku, Veschgini:2010ws, Da_Rold:2005zs, Parnachev:2006dn, Gherghetta:2009ac,
Erdmenger:2007cm, Cai:2007zw, Sachan:2011iy,Andreev:2010bv, Park:2011qq,Cai:2007bq} and spectrum of mesons and baryons are studied in \cite{Zhang:2010tk,Hong:2006ta}.

There are two approaches from where one can construct QCD like theories. These approaches are known as top-down and bottom-up approaches. In top down approach, one starts from stringy D brane configurations and construct models for QCD \cite{Sakai:2004cn, Sakai:2005yt} while in bottom-up approach, one starts from QCD  and attempts to construct its five dimensional gravity dual. These five dimensional dual models can be generalized to study various properties of QCD.  The bottom-up approach is divided into two categories, hard wall\cite{Erlich:2005qh} and soft wall models\cite{Karch:2006pv}. In hard wall model, one imposes a cut-off at IR boundary. The IR cut-off in hard wall model is inverse of the QCD scale. The hard wall model describes many properties of QCD such as form factors, effective coupling constants, chiral symmetry breaking and correlation functions, but fails to accommodate  Regge trajectory of meson masses. The problem of mass spectra can be removed by introduction of a  dilaton field. This model is known as soft wall model of AdS/QCD and  the  IR boundary in this model is shifted to  infinity. 

The transition between confining/deconfining phase is studied by Hawking-Page transition in bulk spacetime \cite{Hawking:1982dh}. The high temperature phase  is charged AdS black hole while low temperature phase is thermally charged AdS geometry. The confinement/deconfinement is studied in hard wall and soft wall models in \cite{Herzog:2006ra} and  models with chemical potential are studied in  \cite{Kim:2007em,Park:2011qq, Sachan:2011iy,Lee:2009bya}.

In the charged black hole solutions, charge of black hole is related to chemical potential of the quarks. The dual gauge theory defining the deconfining  phase is AdS black hole while the confining phase is defined by thermally charged AdS solutions \cite{Lee:2009bya}. The UV divergences in these actions is removed by subtraction of action of thermal AdS  \cite{ Park:2011qq,Lee:2009bya,Cvetic:2001bk}.

The study of gauge/gravity duality provides a relation between the gravity theories in the AdS spacetime and conformal field theories on the boundary of the AdS spacetime. In recent years, a large number of generalized geometries are studied, which gives a dual scale invariant gauge theory.  One of the metric representing such a geometry is given by,
\be
ds^2~=~r^{2\alpha}\left(r^{2z}f(r)~dt^2+\frac{1}{r^2f(r)}dr^2+r^2~d\vec{x}^2\right),\qquad \textrm{where}\qquad \alpha~=~-\theta/d.
\ee
In this metric, the constants $z$ and $\theta$ are dynamical and hyperscaling violation exponents, respectively. This metric gives AdS solutions for $\theta=0$ and $z=1$.
The scale transformations, $t\to \lambda^z,~r\to\lambda^{-1}r,~x_i\to\lambda x_i$
leads to $ds_{d+2}^2\to\lambda^{\theta/d}ds^2_{d+2}$. Thus the transformations retains the spatially homogeneous  and covariant nature, but the distance scales as powers of $\lambda$ for non zero value of $\theta$. The non invariance of distance in reference of AdS/CFT correspondence leads to violation of hyperscaling in dual field theory.  In other words, it is shown in  \cite{Kim:2012pd} if the hyperscaling violation exponent is included in the metric, the entropy scales $T^{(d-\theta)/z}$.  If hyperscaling is not taken into account, the entropy scales as $T^{d/z}$\cite{ Dong:2012se, Kim:2012pd, Gath:2012pg}.
 
In this article, we study the effect of warping on  confinement/deconfinement transition in the simplest case by taking the value of $z=1$. In this context we have only single charge in AdS black hole \cite{Alishahiha:2012qu}. We first study the holographic model of QCD with a dilaton potential in Einstein-Maxwell-Dilaton system. The form of potential is taken to be exponential with some free parameters. These parameters are be fixed by various boundary conditions. We calculate Euclidean actions of charged AdS black hole and thermally charged AdS. These actions are regularized by subtraction of thermal AdS action. The Hawking-Page transition is studied and plots between grand potential vs temperature and transition  temperature vs chemical potential are given. The plots of transition temperature ($T_c$) vs chemical potential $\mu$ in \autoref{fig:tmu1} are plotted  for different values of $\alpha$. The warp factor $\alpha$ depends upon hyperscaling violation exponent, therefore the transition temperature varies with hyperscaling violation exponent. The transition temperature becomes independent of warp factor (or hyperscaling violation exponent) for certain value of $\mu$. This point, where all curves meet may be a signature of second order transition.

 This article is organized in six sections. In \autoref{sec:gs}, we briefly summarized the calculations for $z=1$ and single gauge field. In  \autoref{sec:abh} and  \autoref{sec:tca} , grand potentials are calculated for charged AdS black hole and thermally charged AdS. The \autoref{sec:cdt} is devoted to study of confinement/ deconfinement transition and in  \autoref{sec:con}, we conclude our analysis.

 %%%%%%%%%%%%%%%%%%%%%%%%%%%%%%%%%%%%%%%%%%%%%%%%%%%%%%%%%%%%%%%%%%%%%%%%%%%%%%%%%%%%%%%%%%%%
 \section{Gravitational solution of EMD theory}\label{sec:gs}
 In this section the  solution Einstein-Maxwell-Dilaton system  with hyperscaling violation \cite{Alishahiha:2012qu} is given. We take the solution obtained in \cite{Alishahiha:2012qu} and take  dynamical exponent $z=1$. Taking $z=1$, limits the number of gauge fields in solution to one.  We begin with well known Einstein-Maxwell-Dilaton action with exponential potential ($V=V_0~e^{\gamma\phi}$) given by,
 \be
 S~=~-\int~d^{d+2}x~\sqrt{g}\left[\frac{1}{2\kappa^2}\left(R-\frac{1}{2}(\partial\phi)^2+V_0~e^{\gamma\phi}\right)-\frac{1}{4g^2}F^2e^{\lambda \phi}\right]\label{action}
 \ee
 where $\kappa^2=8\pi G$, $g$ is coupling constant of dimension $d+2$, $\lambda$, $\gamma$ and $V_0$ are  parameters of the model, which will be fixed later.
 
The equations of motion for gravitational part of action \ref{action} can be written as,
\be\label{eom1}
\frac{1}{2\kappa^2}\left\lbrace G_{\mu\nu}-\frac{1}{2}\left(-\frac{g_{\mu\nu}}{2}(\partial \phi)^2+\partial_\mu\phi\partial_\nu\phi\right)-\frac{1}{2}g_{\mu\nu}V_0~e^{\gamma\phi}\right\rbrace
=T_{\mu\nu}
\ee
where
\be
 G_{\mu\nu}~=~R_{\mu\nu}-\frac{1}{2}g_{\mu\nu}R,\qquad
 T_{\mu\nu}~=~\frac{1}{2g^2}e^{\lambda\phi}\left\lbrace F^\rho_{~~\mu}F_{\rho\nu}-\frac{1}{4}F^2g_{\mu\nu}\right\rbrace
\ee
The equations of motion \ref{eom1} can be written in modified form as,
\be
R_{\mu\nu}+\frac{V_0~e^{\gamma\phi}}{d}g_{\mu\nu}~=~\frac{1}{2}\partial_\mu\phi\partial_\nu\phi+\frac{\kappa^2}{g^2}e^{\lambda \phi}\left\lbrace F^\rho_{~~\mu}F_{\rho\nu}-\frac{1}{2d}F^2g_{\mu\nu}\right\rbrace
\ee
The equation of motion for scalar field is,
\be
\frac{1}{\sqrt{g}}\partial_\mu(\sqrt{g}\,g^{\mu\nu}\partial_\nu\phi)~=~-\frac{\partial (V_0~e^{\gamma\phi})}{\partial \phi}+\frac{1}{2}\frac{\kappa^2}{g^2}\lambda e^{\lambda \phi}F^2
\ee
and for gauge field is,
\be
\frac{1}{\sqrt{g}}\partial_\mu(\sqrt{g}\,e^{\lambda\phi}F^{\mu\nu})~=~0.
\ee

Let us consider the ansatz for our metric (with $z=1$), scalar field and gauge field, which are given as,
\be
ds^2~=~r^{2\alpha}\left(r^2f(r)~dt^2+\frac{1}{r^2f(r)}dr^2+r^2~d\vec{x}^2\right),\qquad\phi=\phi(r),\qquad F_{r,t}\neq 0.
\ee
We consider $F_{\mu\nu}$ as only function of $r$ and rest of the components are equal to zero.

Using our ansatz, the solution  for Maxwell's equations can be written as,
\be
F_{rt}~=~e^{-\lambda\phi}r^{\alpha(2-d)-d}\rho,\label{fs1}
\ee
where $\rho$ is integration constant and  to be related to charge of the black hole later.

On solving $tt$ and $rr$ components of Einstein's  equations, we determine the  scalar field, which is given as,
\be
e^\phi~=~e^{\phi_0}r^{\sqrt{2d\alpha (\alpha+1)}}~=~e^{\phi_0}r^\zeta.
\ee
The exponent on $r$  shows that to get well defined solutions, we must have $\alpha (\alpha+1)\geq 0$. 
 
Using equations of motion, the metric function is given by,
\be
f(r)~=~1-\frac{m}{r^{d(1+\alpha)+1}}+\frac{Q^2}{r^{2d(1+\alpha)}}.
\ee
where $m$ is related to the mass of the black hole and   $Q$ is related to $\rho$ by the following relation,
\be
Q^2~=~-\frac{\kappa^2}{g^2}\frac{e^{-\lambda\phi_0}}{d(1+\alpha)(-1+d-2\alpha+d\alpha+\zeta\lambda)}\rho^2.\label{c1}
\ee 
The parameter $\gamma$ appearing in exponential of potential is fixed by using the fact that the  a constant term (independent of $r$) appears in metric.  Equating the powers of $r$ to zero, the parameter $\gamma$ can be fixed as $-2\alpha/\zeta$.  The constant term is equated to unity to get the value $V_0$, which is given by relation,
\be
V_0~=d(1+\alpha)(1+d+d\alpha)e^{-\gamma\phi_0},
\ee
and using equation of motion for scalar field, the value of parameter $\lambda$ is fixed as $-\gamma$.
The solution for field strength \ref{fs1} becomes,
\be
F_{rt}~=~i\bar{Q}~r^{-d(\alpha+1)},
\ee
where we have defined $\bar{Q}=\frac{g}{\kappa}Q\sqrt{d(1+\alpha)(d\alpha+d-1)}~e^{-\lambda\phi_0/2}$. The solution of gauge field $A_t$ is given by,
\be
A_t(r)~=~\frac{i\bar{Q}}{1-d(\alpha+1)}r^{-d(\alpha+1)+1}+C \label{gauge1}
\ee
where $C$ is a constant and related to boundary value of $A_t$, which is chemical potential of the system.

%%%%%%%%%%%%%%%%%%%%%%%%%%%%%%%%%%%%%%%%%%%%%%%%%
\section{AdS Black hole}\label{sec:abh}
In this section, we consider the black hole solution for the warped geometry. The solution of $A_t$ with appropriate boundary conditions leads us to the solution of charged AdS black hole. Using the solution of $A_t$ in \ref{gauge1}, we apply the condition that at the boundary ($r\to \infty$), the  value of $A_t$ is $i\mu$, where $\mu$ is chemical potential of the black hole and $i$ is due to the consideration of Euclidean signature. The boundary value give us the constant $C=i\mu$ and the solution of $A_t$ has the form,
\be
A_t(r)~=~i\left( \mu-\frac{\bar{Q}}{d(\alpha+1)-1}r^{-d(\alpha+1)+1}\right).
\ee
At horizon ($r_H$), $A_t=0$, leads us to the relation between $\mu$ and $\bar{Q}$, which is given by,
\be
\bar{Q}~=~\frac{d(\alpha+1)-1}{r_H^{-d(\alpha+1)+1}}\mu,\qquad\Longrightarrow Q~=~\frac{\kappa}{g}\mu\sqrt{\frac{d(1+\alpha)-1}{d(1+\alpha)}}~e^{\lambda\phi_0/2}~r_H^{d(\alpha+1)-1}
\ee
The radius of horizon for charged black hole solution is obtained by equating the metric function $f(r_H)=0$. This leads to the equation for $r_H$, which is given as,
\be
r_H^{2d(1+\alpha)}-mr_H^{d(1+\alpha)-1}+Q^2~=~0,
\ee
and the Hawking temperature of the black hole given by,
\begin{align}
T~&=~\frac{1}{4\pi}(d\alpha+d+1)~r_H\left(1-Q^2~\frac{d\alpha+d-1}{d\alpha+d+1}~r_H^{-2d(1+\alpha)}\right)\nn\\
&=~\frac{1}{4\pi}(d\alpha+d+1)~r_H\left( 1-\mu^2\frac{\kappa^2}{g^2}\frac{(d\alpha+d-1)^2}{d(1+\alpha)(d\alpha+d+1)}e^{\lambda\phi_0}\frac{1}{r_H^2}\right)\label{haw}.
\end{align}
Now redefining some variables for simplicity,
\be
D_1~=~\frac{d\alpha+d+1}{4\pi}\qquad,\qquad D_2~=~\frac{\kappa^2}{g^2}\frac{(d\alpha+d-1)^2}{d(1+\alpha)(d\alpha+d+1)}e^{\lambda\phi_0}\nn,
\ee
Using these in \ref{haw} and solving the quadratic equation, we get positive value of horizon radius as,
\be
r_H~=~\frac{T+\sqrt{T^2+4\mu^2D_1^2D_2}}{2D_1}.
\ee

Using equation of motion, action \ref{action} can be written as,
\be
S^{AdSBH}~=~\frac{1}{d}\int d^{d+2}x~\sqrt{g}\left[\frac{V}{\kappa^2}+\frac{1}{2g^2}e^{\lambda\phi}F^2\right]~=~\frac{1}{d}V_d\,\beta\int dr~\sqrt{g}\left[\frac{V}{\kappa^2}+\frac{1}{2g^2}e^{\lambda\phi}F^2\right],
\ee
where $V_d$ is $d$ dimensional volume and $\beta$ is inverse of black hole temperature. On substituting various values in the above action, it simplifies to,
\begin{align}
S^{AdSBH}~&=~\frac{1}{d}V_d\,\beta\frac{d(1+\alpha)}{\kappa^2}\int dr~r^{d(\alpha+1)+2\alpha}\left[(d\alpha+d+1)r^{-2\alpha}-Q^2~(d\alpha+d-1)r^{-2d(\alpha+1)-2\alpha}\right]\nn\\
&=~\frac{1}{d}V_d\,\beta\frac{d(1+\alpha)}{\kappa^2}\left[r^{d(\alpha+1)+1}+Q^2~r^{-d(\alpha+1)-1}\right]_{r_H}^{r_{max}}
\end{align}
where we take $r_{max}\to \infty$ at the end of the calculations. 

The above action is singular at $r_{max}\to \infty$. Therefore, to regularize this action we subtract thermal AdS from this action. The metric for thermal AdS is given by,
\be
ds^2~=~r^{2\alpha}\left(r^2~dt^2+\frac{1}{r^2}dr^2+r^2~d\vec{x}^2\right),
\ee
and  action for thermal AdS with time periodicity $\beta_1$ is given by equation,
\be
S^{tAdS}~=~-V_d\,\beta_1\frac{(1+\alpha)}{\kappa^2}(r_{max})^{d(\alpha+1)+1}
\ee
Thus the, regularized action for AdS black hole is given by,
\begin{align}
\bar{S}^{AdSBH}~&=~\lim_{r_{max}\to\infty}V_d\,\beta\frac{(1+\alpha)}{\kappa^2}\left\lbrace\left[r^{d(\alpha+1)+1}+Q^2~r^{-d(\alpha+1)-1}\right]_{r_H}^{r_{max}}\right.\nn\\
&\qquad\qquad\qquad\qquad \qquad\qquad\left.-\left(\frac{f(r_{max})}{f(r_{max},m=Q=0)}\right)^{1/2}r^{d(\alpha+1)+1}\Big\vert_0^{r_{max}}\right\rbrace\nn\\
&=~ -V_d\,\beta\frac{(1+\alpha)}{\kappa^2}\left((r_H)^{d(\alpha+1)+1}+Q^2~(r_H)^{-d(\alpha+1)+1}\right)
\end{align}
The factor $\left(\frac{f(r_{max})}{f(r_{max},m=Q=0)}\right)^{1/2}$ in front of last term in the first expression is inserted to match the Euclidean time periodicity at $r=r_{max}$, where both the solutions coincide with each other. The singular term (powers of $r$ with positive values) of AdS black hole solution is cancelled with the term in thermal AdS solution and we get the regularized action. By using  thermodynamical relation, $\Omega(\mu ,T)=T\,S_{on-shell}$, we write  the regularized grand potential for AdS black hole as,
\be
\Omega^{AdSBH}~=~-V_d\frac{1+\alpha}{\kappa^2}\left((r_H)^{d(\alpha+1)+1}+(r_H)^{-d(\alpha+1)+1}Q^2\right)
\ee

%The susceptibility is defined by the relation, $\chi(T)=-\frac{1}{V_d}\frac{\partial^2\Omega(\mu,T)}{\partial\mu^2}$, thus susceptibility in this phase is given by,
%\be
%\chi^{AdSBH}(T)~=~
%\ee
%%%%%%%%%%%%%%%%%%%%%%%%%%%%%%%%%%%%%%%%%%%%%%%%%%%%%
\section{Thermally charged AdS}\label{sec:tca}
This section is devoted to the study of thermally charged AdS solution \cite{Lee:2009bya}. The thermally charged AdS is also asymptotically AdS but doesn't have a horizon. Due to absence of horizon, we choose a lower cut-off for thermally charged AdS as $r_{IR}$ and integrate from $r_{IR}$ to $\infty$. The metric function for thermally charged AdS is given by,
\be 
f_1(r)~=~1+\frac{Q_1^2}{r^{2d(\alpha+1)}}.
\ee
where $Q_1$ is charge associated with thermally charged AdS . This metric function also satisfies Einstein-Maxwell equations. This geometry is simply obtained by putting $m=0$ in solution of AdS black hole. The charge $Q_1$ in this case is different from that of  AdS black hole due to different boundary conditions. 

The field strength tensor for thermally charged AdS is given by same equation as that for AdS black hole case, but now $Q$ is replaced by $Q_1$. Th expression is written as,
\be 
F_{1rt}~=~i\bar{Q}_1\,r^{-d(\alpha+1)},\qquad\textrm{where}~ \bar{Q}_1=\frac{g}{\kappa}Q_1\sqrt{d(\alpha+1)(d\alpha+d-1)}~e^{-\lambda\phi_0/2}.
\ee
From this field strength, the gauge field can be calculated as, 
\[
A_{1t}(r)~=~\frac{i\bar{Q}_1}{1-d(\alpha+1)}r^{-d(\alpha+1)+1}+C_2.
\]
Again at $r\to\infty$, we have $ A_{1t}(\infty)=C_2=i\mu$, but at $r=r_{IR}$, we apply Dirichlet boundary condition $A_{1t}(r_{IR})=i\xi\mu$, where $\xi$ is a constant to determined. Thus, at $r_{IR}$,
\begin{align}
A_{1t}(r_{IR})~&=~i\xi\mu~=~i\mu-\frac{i\bar{Q}_1}{d(\alpha+1)-1}r^{-d(\alpha+1)+1}\nn\\
\implies&~Q_1~=~\frac{\kappa}{g}~\mu~(1-\xi)~\sqrt{\frac{d(\alpha+1)-1}{d(\alpha+1)}}~e^{\lambda\phi_0/2}~r_{IR}^{d(\alpha+1)-1}
\end{align}

Using same procedure as done for AdS black hole, we compute regularized action for thermally charged AdS, which is written as,
\begin{align}
\bar{S}^{tcAdS}~&=~\lim_{r_{max}\to\infty}V_d\,\beta_1\frac{(1+\alpha)}{\kappa^2}\left\lbrace\left[r^{d(\alpha+1)+1}+Q_1^2~r^{-d(\alpha+1)-1}\right]_{r_{IR}}^{r_{max}}\right.\nn\\
&\qquad\qquad\qquad\qquad \qquad\qquad\left.-\left(\frac{f_1(r_{max})}{f_1(r_{max},Q_1=0)}\right)^{1/2}r^{d(\alpha+1)+1}\Big\vert_0^{r_{max}}\right\rbrace\nn\\
&=~ -V_d\,\beta_1\frac{(1+\alpha)}{\kappa^2}\left((r_{IR})^{d(\alpha+1)+1}+Q_1^2~(r_{IR})^{-d(\alpha+1)+1}\right),
\end{align}
and grand potential for thermally charged AdS is given by equation,
\be
\Omega^{tcAdS}~=~-V_d\frac{1+\alpha}{\kappa^2}\left((r_{IR})^{d(\alpha+1)+1}+Q_1^2~(r_{IR})^{-d(\alpha+1)+1}\right),
\ee
where $Q_1$ is function of chemical potential $\mu$. Using thermodynamical relation $N=-\partial\Omega/\partial\mu$, we calculate the quark number for thermally charged AdS, which is given by relation,
\be
N~=~2\frac{(\alpha+1)}{d~g^2}\,\mu\,(1-\xi)^2\left(d(\alpha+1)-1\right)e^{\lambda\phi_0}~r_{IR}^{d(\alpha+1)-1}.\label{ndensity}
\ee
As shown in \cite{Lee:2009bya}, one has to use Dirichlet boundary condition instead of Neumann, we get free energy from Legendre transformation of  Grand potential and $\mu N \beta_1$ is equal to boundary action $S^{tcAdS}$. Calculating the boundary action, we can determine unknown parameter $\xi$. The boundary action of thermally charged AdS is given by,
\begin{align}
S^{tcAdS}_b~&=~\frac{1}{d}\int_{\partial M}d^{d+1}x~\sqrt{g^{(d+1)}}~\eta^\sigma A_\rho~ F_{\mu\sigma}~g^{\rho\mu}~e^{\lambda\phi}\nn\\
&=~\frac{1}{d~g^2}V_d~\beta_1~\mu~\bar{Q}_1~e^{\lambda\phi}\label{boundary}
\end{align}
where unit vector $\eta^r=(0,-\sqrt{f_1(r)}/r^{\alpha-1},0,0\ldots)$ and $g^{d+1}=r^{(\alpha+1)(d+1)}\sqrt{f_1(r)}$. Comparing \ref{ndensity} and \ref{boundary}, we evaluated constant $\xi=1/2$. Thus charge $Q_1$ of thermally charged AdS is given by,
\be
Q_1~=~\frac{\kappa}{2g}~\mu~\sqrt{\frac{d(\alpha+1)-1}{d(\alpha+1)}}~e^{\lambda\phi_0/2}~r_{IR}^{d(\alpha+1)-1}.
\ee
%The susceptibility in this phase is,
%\be
%\chi^{AdSBH}(T)~=~\frac{9 (d\,\alpha+d-1)  (r_{IR})^{ d\,\alpha+d-1}}{2 d\,g^2}e^{\lambda  %\phi_0 }
%\ee
%%%%%%%%%%%%%%%%%%%%%%%%%%%%%%%%%%%%%%%%%%%%%%%%
\section{Confinement/deconfinement transition}\label{sec:cdt}
\begin{figure}[!htp]
    \centering
    \subfloat[For $\alpha=0.1$\label{fig:grand1}] {%
      \includegraphics[scale=.9]{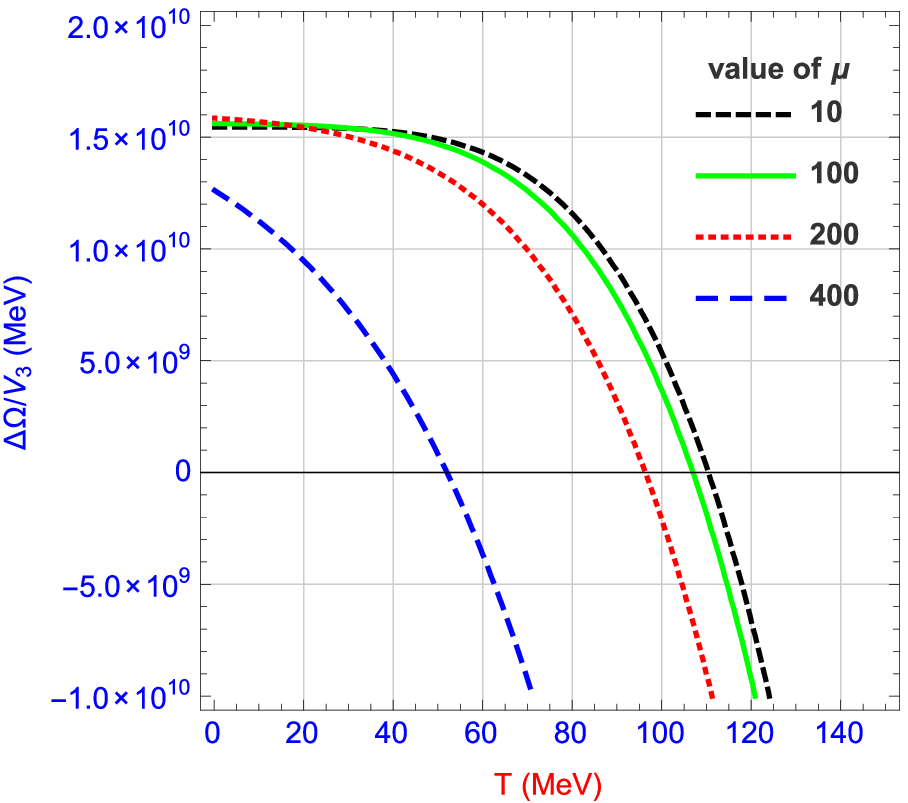}
    }%
    ~~~\subfloat[For $\alpha=0.3$\label{fig:grand2}] {%
      \includegraphics[scale=.9]{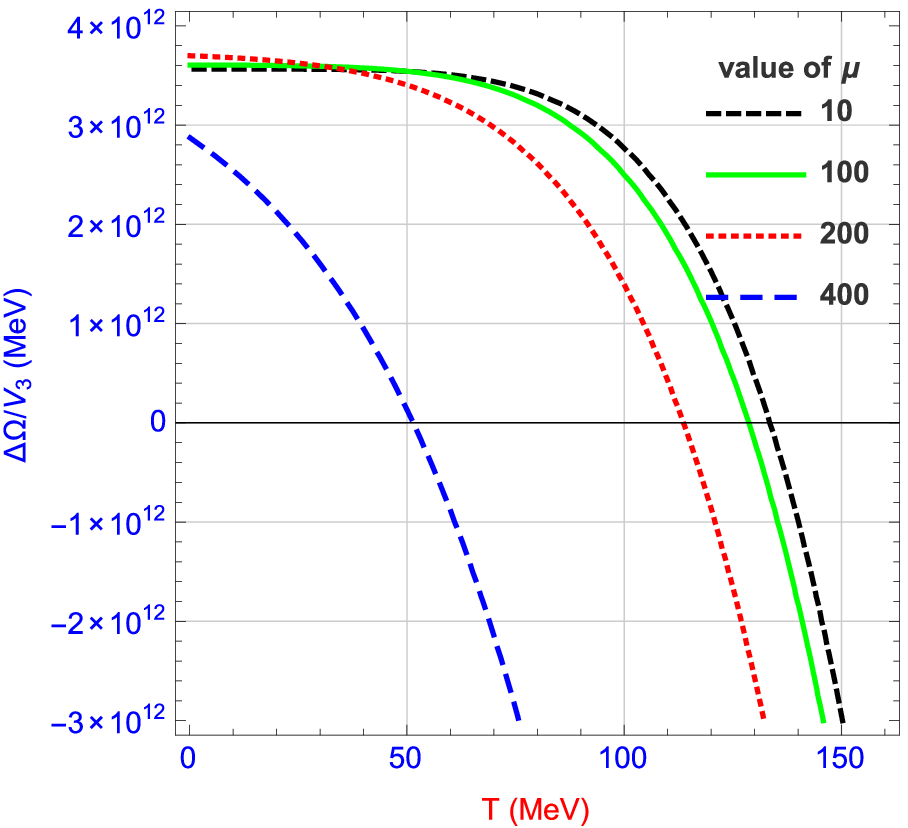}
    } 
    \caption{Grand potential vs temperature at various values of $\mu$ for constant $\alpha$  .}
    \label{fig:grand}
  \end{figure}

Now we study the transition from AdS black hole phase to thermally charged AdS. To study this, we take the difference between the actions of AdS black hole and thermally charged AdS geometries with appropriate periodicity matching. The difference in grand potentials is proportional to difference in actions. The difference in actions is given by,
\begin{align}
\Delta S~&=~\lim_{r_{max}\to\infty}\frac{1}{d}V_d\,\beta\frac{d(1+\alpha)}{\kappa^2}\left\lbrace\left[r^{d(\alpha+1)+1}+Q^2~r^{-d(\alpha+1)-1}\right]_{r_H}^{r_{max}}\right.\nn\\
&\qquad \qquad\qquad\left.-\left(\frac{f(r_{max})}{f_1(r_{max})}\right)^{1/2}\left[r^{d(\alpha+1)+1}+Q_1^2~r^{-d(\alpha+1)-1}\right]_{r_{IR}}^{r_{max}}\right\rbrace\nn\\
&=~ V_d\,\beta\frac{1}{\kappa^2}\left\lbrace\left((r_{IR})^{d(\alpha+1)+1}-(r_H)^{d(\alpha+1)+1}\right)\right.\nn\\
&~~~~~~~+\left.\mu^2\{d(1+\alpha)-1\}~e^{\lambda\phi_0}\left(\frac{1}{4}~r_{IR}^{d(\alpha+1)-1}-r_H^{d(\alpha+1)-1}\right)\right\rbrace.
\end{align}
The factor $\left(\frac{f(r_{max})}{f_1(r_{max})}\right)^{1/2}$ in front of second term  comes from periodicity matching of AdS black hole and thermally charged AdS geometries. Using this expression, we calculated grand potential, which is given as,
\begin{align}
\Delta\Omega~=&~ V_d\,\frac{1}{\kappa^2}\left\lbrace\left((r_{IR})^{d(\alpha+1)+1}-(r_H)^{d(\alpha+1)+1}\right)\right.\nn\\
&~~~~~~~+\left.\mu^2\{d(1+\alpha)-1\}~e^{\lambda\phi_0}\left(\frac{1}{4}~r_{IR}^{d(\alpha+1)-1}-r_H^{d(\alpha+1)-1}\right)\right\rbrace.
\end{align}
If the value of $\Delta\Omega$ is less than zero, the dominant geometry is AdS black hole and vice-versa. The sign of $\Delta\Omega$ governs the nature of  the stability of phase and confinement/deconfinement  transition. The value of $r_{IR}$ is set to be $323~MeV$, which is calculated from the mass of lightest mesons \cite{Herzog:2006ra}.
We have plotted grand potential  difference ($\Delta\Omega$) vs temperature ($T$) for various values of warp factor $\alpha$ which are given in \autoref{fig:grand}.

The relation describing the  five dimensional gravitational  constant and that of the five dimensional gauge coupling  constant are evaluated by the application of AdS/CFT to QCD. These constants relates the colour gauge group ($N_c$) and number of flavours ($N_f$) as,
\begin{equation}
\frac{1}{2\kappa^2}~=~\frac{N_c^2}{8\pi^2}~~\textrm{and}~~\frac{1}{2g^2}~=~\frac{N_cN_f}{8\pi^2}
\end{equation}
In our study, we have used $N_f=2$ and $N_c=3$.\\
%%%%%%%%%%%%%%%%%%%%%%%%%%%%%%%%%%%%%
The value of $\alpha$ is considered to be equal to $-\theta/d$, which is commonly used in literature mentioned in \cite{Alishahiha:2012qu}. We have taken dimension for our estimation to be five, i.e $d=3$ and $\phi_0=0$. To get plot $T$ vs $\mu$, we equate $\Delta\Omega=0$ and these plots for various values of $\alpha $ are given in \autoref{fig:tmu1}.
%%%%%%%%%%%%%%%%%%%%%%%%%%%%%%%%%%%%%%%%%%
\section{Conclusion}\label{sec:con}
In this article , we studied the thermodynamic behavior of AdS/QCD from holographic approach with generalized warp factor.
The plots of grand potential per unit volume are shown in \autoref{fig:grand}. The figure \autoref{fig:grand1} shows grand potential per unit volume vs temperature for $\alpha=0.1$ and figure \autoref{fig:grand2} for $\alpha=0.3$. These plots shows that the increasing value of chemical potential $\mu$, for constant $\alpha$, transition temperature got lowered but the maximum value of grand potential increases, which indicates the stability of thermally charged AdS at lower temperatures.
\begin{figure}
  \begin{center}
    \includegraphics[height=0.5\textwidth,width=0.9\textwidth]{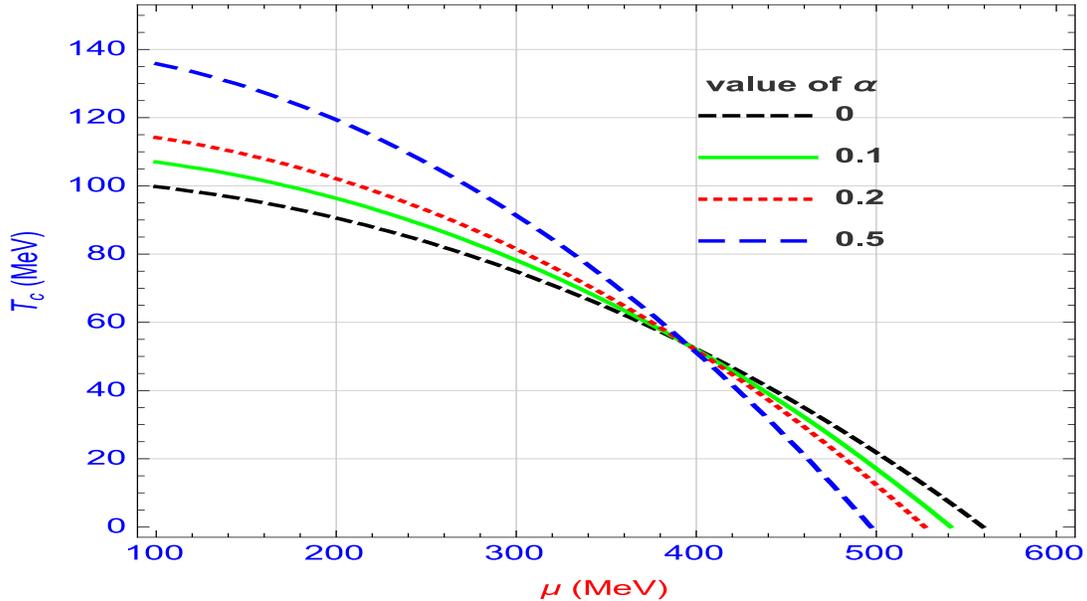}
    \caption{$T$ vs $\mu$ for various values of  $\alpha$.}
    \label{fig:tmu1}
  \end{center}
  \vspace{-20pt}
 % \vspace{1pt}
\end{figure} 
The entropy difference  $\Delta S$, (entropy is defined as $ S=-(\partial\Omega/\partial T)_{V,\mu}$) is non zero, which shows the transition is of first order (using Ehrenfest  scheme for classification of phase transition). 

The \autoref{fig:tmu1}, shows the plot between chemical potential and transition temperature for various values of $\alpha$. The results obtained here shows similar qualitative  behavior with various results obtained without warping dependence on dimension except the fact that for different values of warping, all plots of \autoref{fig:tmu1} meet at a point. It means that the transition is independent of warping on this point. We believe that it is the onset of second order transition. This is also expected from recent lattice data \cite{Fromm:2011qi}. It would be interesting to study mass spectra of mesons in this scenario, transport properties and corrections arising due to  Gauss-Bonnet gravity.

\section*{Acknowledgments:}
S. Sachan is supported by CSIR-Senior Research
Fellowship, grant no. (09/013(0239)/2009-EMR-I).  I would also like to thank  Dr. Sanjay Siwach for discussing the problem at various stages of this work. 
%\clearpage
%%%%%%%%%%%%%%%%%%%%%%%%%%%%%%%%%%%%%%%%%%%%%%%%%%%%%%%%%%%%%%%%%%%%%%%%%%%%%%%%

\end{document}